\documentclass[aps,prd,amsfonts,amsmath,amssymb,showkeys,showpacs,a4paper,preprint,tightenlines]{revtex4-1}
\usepackage{graphicx}
\usepackage{subfig}
\usepackage{bm}
\usepackage{ulem}
\usepackage{dcolumn}
\usepackage{hyperref}

\begin{document}

\title{Testing the tensor-vector-scalar Theory with the
latest cosmological observations}
\author{Xiao-dong Xu, Bin Wang, Pengjie Zhang}

\affiliation{IFSA Collaborative Innovation Center, Department of Physics and Astronomy, Shanghai Jiao Tong University, Shanghai 200240, China}

\begin{abstract}

The tensor-vector-scalar (TeVeS) model is considered a
viable theory of gravity. It produces the Milgrom's
modified Newtonian dynamics in the
nonrelativistic weak field limit and is free from
ghosts.  This model has been tested against various cosmological observations. Here we investigate whether new observations such as the galaxy velocity power spectrum measured by 6dF and the kinetic Sunyaev Zel'dovich effect power spectrum measured by ACT/SPT can put further constraints on the TeVeS model. Furthermore, we perform the test of TeVeS cosmology with a sterile neutrino by confronting to {\it Planck} data, and find  that it is ruled out by cosmic microwave background measurements from the {\it Planck} mission.

\end{abstract}



\maketitle

\section{Introduction}

The convincing observational evidences from the
scale of galaxies to the scale of the cosmic
microwave background (CMB) radiation accumulated
over the past few decades raised the missing mass problem: there is a mismatch between the
dynamics and distribution of visible matter
\cite{Oort1932, Zwicky1933, Smith1936, Rubin1970, Rubin1980, Schmidt1998, Perlmutter1999, Sofue2001, Clowe2006, Nolta2009}.
To explain this problem,
one usually postulates the existence of a new form
of matter in nature, called dark matter (DM).
DM is considered nonbaryonic and does not
emit light or interact with electromagnetic
field. For now people only detect DM through
its gravitational effect. Traditionally, DM can
be classified as ``hot dark matter,'' which is
composed of relativistic particles such as
massive neutrinos; ``cold dark matter'' (CDM), which is
composed of very massive slowly moving and weakly
interacting particles; and in between the
possible ``warm dark matter,'' which is also
sometimes considered. One attributes the observed
extra gravitational force to the DM component whose
abundance greatly exceeds the
visible matter. In the standard
$\Lambda$CDM model, DM contributes about 25\% to
the total energy budget in the Universe. The discrepancy between the dynamics and
distribution of visible matter happens on galactic
to cosmological scales. Decades after the
proposal of DM, it was discovered that the
expansion of our Universe is accelerating, which
calls for another new substance, dark energy (DE),
to contribute the mysterious missing energy at
cosmological scales.

Einstein's general
relativity (GR) has been vigorously tested in the
Solar System, but on galaxy or larger scales its
validity has not been completely proved.
Considering that the law of gravity plays a
fundamental role at every instance where
discrepancies have been observed, it is possible
that the phenomena attributed to DM and DE are
just a different theory of gravity in disguise.
The research relating to modifications of gravity theory
is not extensive. In the literature, modified
gravity theories usually contain a Newtonian
limit for the low velocity, weak potential case.
Considering that the mass discrepancy problem appears
on extragalactic scales where Newtonian gravity
is expected to be a good approximation, these
theories cannot solve the problem without the
help of the invisible matter component. This has been
resolved in the Milgrom's modified Newtonian
dynamics(MOND) proposal\cite{Milgrom1983,
Milgrom1983a, Milgrom1983b}, which assumes that
Newtonian gravity fails in low acceleration
cases. Instead, the acceleration $a$ induced by
the gravitational force was proposed as $
\tilde{\mu}(a/a_0)a = -\nabla \Phi_{\mathrm{N}},
$ where $a_0$ is a characteristic acceleration
scale, and $\Phi_{\mathrm{N}}$ is the usual Newtonian
potential. $\tilde{\mu}(x) \simeq x$ for $x \ll
1$ and $ \tilde{\mu}(x) \rightarrow 1$ for $x \gg
1$. In laboratory and solar system experiments,
$a \gg a_0$, MOND returns to the Newtonian
dynamics; while in the extragalactic regime where $a
\ll a_0$, the acceleration squared is
proportional to the gravitational force. MOND is
extremely successful in explaining galactic
rotation curve\cite{Milgrom1988, Begeman1991,
Sanders1996, Blok1998, McGaugh1998, Milgrom2005,
Milgrom2007} and the Tully-Fisher law
\cite{McGaugh2000, McGaugh2005}. Some other
predictions of MOND can be found
in\cite{Sanders1990, Milgrom1998, Sanders2002,
Skordis2009, Cardone2010, Famaey2011}.

To be able to make predictions for cosmological
observations, a relativistic theory of MOND is
required.  After some early
attempts\cite{Bekenstein1984, Bekenstein1988,
Sanders1988, Bekenstein1994, Sanders1997},
Bekenstein succeeded in constructing
tensor-vector-scalar (TeVeS)
theory\cite{Bekenstein2004}, which is a
relativistic theory of gravity and produces
MOND in the nonrelativistic weak field limit. The
name comes from the fact that the theory contains a scalar
and a vector field in addition to the metric (a
tensor field). TeVeS theory has proven
successful in explaining the astrophysical data
at scales larger than that of the Solar System
without the need of an excessive amount of
invisible matter \cite{Ferreras2008,
Ferreras2009, Mavromatos2009, Chiu2006, Chen2006,
Shan2008, Chen2008, Feix2010, Chiu2011}. Moreover, TeVeS
theory has proven to be free of
ghosts\cite{Chaichian2014}, which makes TeVeS,
including its nonrelativistic limit, a viable
theory of gravity.

In order to predict large scale structure
observations in TeVeS theory, we need the
linear cosmological perturbation theory in TeVeS,
which was constructed in a pioneer work\cite{Skordis2006}. Based on the perturbation
theory, the large scale structure in TeVeS
cosmology was first discussed in \cite{Skordis2006a}, where it was argued that
perturbations of the scalar field may induce
enhanced growth in the matter perturbations.
Analytic explanation of the growth of structure
was subsequently given in \cite{Dodelson2006},
where it was claimed that the perturbations of the vector field are key to the enhanced growth. It was
further clarified in \cite{Skordis2009} that even
if the contribution of the TeVeS fields to
the background Friedmann equations is negligible, one
can still get a growing mode that drives
structure formation. This explains analytically
the numerical results in \cite{Skordis2006a}.

It is of great interest to examine whether TeVeS
theory can give predictions for large scale
structure similar to the $\Lambda$CDM model and whether it is
compatible with cosmological observations. In \cite{Reyes2010}, Reyes {\it et al.} reported the measurement of $E_G$, an estimator of the ratio of the Laplacian of gravitational potential to the peculiar velocity divergence\cite{Zhang2007}, using a sample of 70,205 luminous red galaxies in the redshift range [0.16, 0.47] from the Sloan Digital Sky Survey. They claimed that the original Bekenstein's TeVeS model is excluded at $2.5\sigma$. Since $E_G$ measures the ratio of two types of perturbations, it is insensitive to the overall amplitude of perturbation.  Other than $E_G$, observations such as the galaxy velocity power spectrum and the kinetic Sunyaev Zel'dovich (kSZ) effect are sensitive to the perturbation amplitude. It is intriguing to investigate whether these probes can put complementary constraints on TeVeS and modified gravity theories in general. This motivates us to further test TeVeS against these complementary observations in large scale structure and examine whether these tests can distinguish TeVeS from $\Lambda$CDM, which serves as the first motivation of the paper.

The mechanism of structure growth in TeVeS theory
is different from that in the $\Lambda$CDM model.
In $\Lambda$CDM, after decoupling from photons,
baryons fall into the gravitational wells induced
by CDM. While in TeVeS, the growth of
perturbations is driven by the vector field whose perturbation grows rapidly after
recombination\cite{Dodelson2006}. This may lead to difference in the growth of
baryon density perturbation and the amplitude of the
matter peculiar velocity. The change
on the matter peculiar velocity can further induce
temperature fluctuations on the CMB map at small
scales via the conventional kSZ effect. The kSZ effect is generated through CMB photons scattering off free electrons in the diffuse
intergalactic medium and the unresolved cluster
population. The study of the kSZ effect is
appealing, since it can be observed with the new
generation CMB experiments. Recently, the kSZ
effect has been found as a potential probe of
reionization, the radial inhomogeneities in the
Lemaitre-Tolman-Bondi cosmology\cite{Bull2011},
the missing baryon problem\cite{Genova-Santos2009}, the dark flow\cite{Zhang2010} and the interaction between the
dark sectors\cite{Xu2013}. Here, we further
investigate the kSZ effect in the frame of TeVeS theory, and disclose whether it can be used to constrain the TeVeS model.

In addition to the signatures in the kSZ effect, we also consider the growth rate
of baryon density perturbation in TeVeS theory.
The growth rate is generally a function of the
cosmic scale factor $a$ and the comoving
wave number $k$, defined as $f(k;a)= d \ln
\delta(k;a)/d \ln a$. Although the temporal
dependence of the growth rate has been readily
measured by galaxy surveys using redshift-space
distortion measurements\cite{Beutler2013,
Blake2011, Torre2013}, its spatial dependence is
currently only weakly constrained\cite{Bean2010,
Daniel2013, Johnson2014}. However the theoretical
study of the latter has undoubted importance,
for it is a critical test of theories of gravity.
A characteristic prediction of $\Lambda$CDM is a
scale-independent growth rate, while modified
gravity models commonly induce a scale dependence
in the growth rate. Thus the measurement of the
growth rate, especially its spatial dependence
can distinguish modified gravity theories from
the standard $\Lambda$CDM model, even if they
produce the same expansion history of the
Universe. In this work we examine the scale
dependence of growth rate in TeVeS and see
whether TeVeS can be distinguished from the $\Lambda$CDM
model using current observations. Since changes in the density/velocity growth rate and scale dependence of the growth rate are generically expected in modified gravity models, the tests we carry out for the TeVeS model can, in principle, be applied to other modified gravity models. Our study on the TeVeS model here then serves as an example to demonstrate possible impacts of these new observations on tests of general relativity at cosmological scales.

The observational data of the probes we proposed above suffer large uncertainties in present observations. Thus, in order to put a tight constraint on the TeVeS model with current data, it is necessary to confront the model with other complementary observations on different scales and redshifts whose precise measurements are already available.  For this purpose we extend our study of the TeVeS cosmology to the CMB since the most accurate observational data on cosmological scales to date come from the CMB experiments. In \cite{Skordis2006a} the CMB angular power
spectrum for the TeVeS was first calculated
numerically by solving the linear Boltzmann
equations in the case of TeVeS theory. By using
the initial conditions close to adiabatic, it was
found that the power spectrum provides poor fit to observations compared to the $\Lambda$CDM model. It was
observed that if a cosmological constant and/or
three massive neutrinos are incorporated into the
matter budget, the first peak of the CMB angular
power spectrum could be located at the right
position\cite{Skordis2006a}. Later it was argued that by including a fourth
sterile neutrino, a MOND-like theory
can have good fits to the CMB angular power
spectrum\cite{Angus2008}. However, in this
research, it was assumed that there were no MOND
effects before recombination, so that the MOND effects do not influence the CMB power spectrum. Thus, it
would be fair to say that their fitting result
has nothing to do with TeVeS features. In this
work, we take into account the full TeVeS features
and their corresponding influences on the CMB. We examine whether we can get a good fit to current
CMB observations by including the
cosmological constant and the fourth neutrino. Considering the high precision of {\it Planck} results, we expect that the CMB observations can give tight constraints on TeVeS cosmology.

The paper is organized as follows. In Sec.\ref{sec.fundamental}, we
go over the TeVeS model and its application
in cosmology. In the
following section, we examine the evolution of
the density perturbation (Sec.\ref{sec.growth}) and the baryon peculiar
velocity (Sec.\ref{sec.velocity}) in TeVeS theory. In Sec.\ref{sec.ksz} we show that the
kSZ effect is a potential probe to constrain the
TeVeS model. In Sec.\ref{sec.scaleDepend}, we
focus on the scale dependence of the growth rate in the TeVeS model and compare with that of the
$\Lambda$CDM model and observational data. In Sec.\ref{sec.cmb}, we concentrate on its influence on
the CMB angular power spectrum in the presence of the sterile neutrino and we confront the TeVeS model with {\it Planck}
data. Finally we draw the conclusions in Sec.\ref{sec.conclude}.

\section{Fundamentals of TeVeS Theory}\label{sec.fundamental}

There are two metrics in Bekenstein's TeVeS theory\cite{Bekenstein2004}. In
addition to the Einstein frame metric $\tilde{g}_{\mu\nu}$ whose
dynamics is governed by the standard
Einstein-Hilbert action, it also has the matter frame
metric $g_{\mu\nu}$. These two metrics are related
through\cite{Bekenstein2004}
\begin{equation}
g_{\mu\nu} = e^{-2\phi} \tilde{g}_{\mu\nu} - 2
\sinh(2\phi) A_{\mu} A_{\nu},
\end{equation}
where $\phi$ is a scalar field and $A_\mu$ is a vector field. The vector field is required to be unit
timelike in the Einstein frame, $
\tilde{g}^{\mu\nu} A_{\mu} A_{\nu} = -1$. The dynamics of the
scalar and vector fields is given by the action $S_{\phi}$ and $S_A$:
\begin{equation}
S_{\phi} = -\frac{1}{16\pi G} \int \mathrm{d}^4 x
\sqrt{-\tilde{g}} [\mu (\tilde{g}^{\mu\nu} -
A^{\mu}A^{\nu}) \tilde{\nabla}_{\mu}\phi
\tilde{\nabla}_{\nu}\phi + V(\mu)],
\end{equation}
\begin{equation}
S_A = -\frac{1}{32\pi G} \int \mathrm{d}^4 x
\sqrt{-\tilde{g}} [K_B F_{\mu\nu} F^{\mu\nu} -
2\lambda(A_{\mu} A^{\mu} + 1)],
\end{equation}
where $\mu$ is a nondynamical dimensionless
scalar field, $F_{\mu\nu} \equiv
2\tilde{\nabla}_{[\mu} A_{\nu]}$, $F^{\mu\nu} =
\tilde{g}^{\mu\alpha} \tilde{g}^{\nu\beta}
F_{\alpha\beta}$, $A^{\mu} = \tilde{g}^{\mu\nu}
A_{\nu}$,  $\lambda$ is a Lagrange multiplier
ensuring the unit timelike constraint on
$A_{\mu}$ and $K_B$ is a dimensionless constant.
$G$ is the bare gravitational constant, whose
value does not equal to the measured Newton's
constant. The relation between the gravitational constant and Newton's constant depends on
the quasistatic, spherically symmetric solution
to the TeVeS field equations and the free
function $V(\mu)$ \cite{Bekenstein2004, Giannios2005, Bourliot2007, Sagi2008}.
$V(\mu)$ typically depends on a scale $l_B$. In Bekenstein's original work, he proposed\cite{Bekenstein2004}
\begin{equation}
\frac{\mathrm{d}V}{\mathrm{d}\mu} = -\frac{3}{32\pi l_B^2 \mu_0^2} \frac{\mu^2(\mu-2\mu_0)^2}{\mu_0-\mu}, \label{eq.VBek}
\end{equation}
where $\mu_0$ is a dimensionless constant. A
generalization to this function was proposed in
\cite{Bourliot2007}. Sanders\cite{Sanders2006}
and Angus {\it et al.} \cite{Angus2006} suggested
alternative functions that also lead to MOND.

The action for matter fields is usually written
in the matter frame, where it takes the same form
as in GR. Hence the matter frame metric is
sometimes called the physical metric. Generically
denoting the matter fields by $\chi^A$, we have
\begin{equation}
S_m = \int \mathrm{d}^4 x \sqrt{-g} \mathcal{L}[g, \chi^A, \partial\chi^A].
\end{equation}

\subsection{Background dynamics in TeVeS cosmology}

The solutions for the homogeneous and isotropic
universe in TeVeS theory have been studied in
\cite{Bekenstein2004, Diaz-Rivera2006, Bourliot2007, Ferreira2008, Zhao2007, Hao2009}.
Assuming that the spacetime is flat, the physical
metric takes the form
\begin{equation}
\mathrm{d}s^2 = a^2 (-\mathrm{d}\eta ^2 +
\mathrm{d}r^2),
\end{equation}
and the Einstein metric has the similar form
\begin{equation}
\mathrm{d}\tilde{s}^2 = b^2
(-e^{-4\phi}\mathrm{d}\tilde{\eta}^2 + \mathrm{d}\tilde{r}^2 ).
\end{equation}
$a$ and $b$ are the scale factors in the matter
and Einstein frames. They are related through
$a=be^{-\phi}$. In the Einstein frame, the
Friedmann equation reads\cite{Skordis2006}:
\begin{equation}
3\frac{\dot{b}^2}{b^2}=a^2 \bigg[ \frac{1}{2}e^{-2\phi}(\mu
V'+V)+8\pi G e^{-4\phi}\rho \bigg],
\end{equation}
where $\rho$ is the matter energy density that does not
include the scalar field. The vector field is not dynamical in FLRW
cosmology. It always points to the time
direction, and does not contribute to the total
energy density. The background dynamics is completely
described if we have the equation of
motion for $\phi$,
\begin{equation}
\ddot{\phi} = \dot{\phi} \bigg( \frac{\dot{a}}{a}-\dot{\phi} \bigg) - \frac{1}{U} \bigg[ 3\mu\frac{\dot{b}}{b}\dot{\phi} + 4\pi Ga^2e^{-4\phi}(\rho+3P) \bigg],
\end{equation}
where $U \equiv \mu+2V'/V''$ and $P$ denotes the
pressure that does not include the pressure of
the scalar field.

In the matter frame, the Hubble parameter is
defined as $H \equiv \frac{\dot{a}}{a^2}$, where
the dot denotes the derivative with respect to the
conformal time in the matter frame. The
effective Friedmann equation then
reads\cite{Skordis2006}
\begin{equation}
3H^2 = 8\pi G_{\mathrm{eff}} (\rho +
\rho_{\phi}), \label{eq.Friedmann}
\end{equation}
where the effective gravitational constant is $
G_{\mathrm{eff}} = G \frac{e^{-4\phi}}{(1 +
\frac{\mathrm{d}\phi}{\mathrm{d}\ln{a}})^2}. $
The effective energy density of the scalar field is
\begin{equation}
\rho_{\phi} = \frac{1}{16\pi G} e^{2\phi} (\mu V' + V).
\end{equation}

If the free function $V$ takes the form of
(\ref{eq.VBek}), the scalar field energy density will
track the matter energy density\cite{Dodelson2006, Skordis2006a, Skordis2009}. Defining the
effective density fraction as $\Omega_{\phi} = \frac{\rho_{\phi}}{\rho + \rho_{\phi}}$, the
tracker is $ \Omega_{\phi} = \frac{(1+3w)^2}{6(1-w)^2 \mu_0}, $ where $w$ is
the equation of state of the background matter
field. The typical value of $\mu_0$ has the order
of $10^2$, so the scalar field is always
subdominant in the history of our Universe.

We are free to add an arbitrary integration
constant to $V$. This will only change the
Lagrangian of the scalar field by a constant,
thus has no influence on the field equations and
the evolution of the gravitational fields. Adding
a constant in $V$ is equivalent to include a
cosmological constant in the effective Friedmann equation
(\ref{eq.Friedmann}). This leads to the desired
accelerated expansion of our Universe.

\subsection{Linear perturbation theory in TeVeS cosmology}

In this subsection we go over the linear
perturbation theory on the background described above.
This will allow us to link TeVeS theory with
observations of structure formation on large scale
as well as the CMB anisotropies.

The linear perturbation theory for TeVeS
cosmology was first constructed in
\cite{Skordis2006}. We employ the formalism
in \cite{Skordis2006} and consider only scalar
perturbations.

We work under the conformal synchronous gauge, for which $\delta g_{00} = \delta g_{0i} = 0$ and $\delta g_{ij} = 2H_L\delta_{ij} + (\partial_i\partial_j - \frac{1}{3}\delta_{ij}\Delta)H_T$. It is conventional to write in Fourier space $H_L = h/6$ and $H_T = -(h+6\eta)/k^2$. The evolution equations for the matter density
contrast and velocity take the same forms as GR
in the matter frame
\begin{align}
\dot{\delta} =& -3\frac{\dot{a}}{a}(C_s^2-w)\delta - (1+w)\bigg(k^2\theta + \frac{1}{2}\dot{h}\bigg), \label{eq.Ddelta}\\
\dot{\theta} =& -\frac{\dot{a}}{a}(1-3w)\theta + \frac{C_s^2}{1+w}\delta - \frac{\dot{w}}{1+w}\theta - \frac{2}{3}k^2\Sigma.
\end{align}

We denote the perturbation to the scalar field by
$\varphi$, so that $\phi = \bar{\phi} + \varphi$. The vector field perturbation is defined as $A_{\mu} \equiv \bar{A}_{\mu} + ae^{-\bar{\phi}}\alpha_{\mu}$. Its scalar mode is
$\Delta\alpha = \nabla\cdot\vec{\alpha}$. The
evolution equations for the scalar field are given by
\begin{align}
\dot{\varphi} =& -\frac{1}{2U}ae^{-\bar{\phi}}\gamma - \dot{\bar{\phi}}\varphi, \\
\begin{split}
\dot{\gamma} =& -3\frac{\dot{b}}{b}\gamma + \frac{\bar{\mu}}{a}e^{-3\bar{\phi}}k^2(\varphi + \dot{\bar{\phi}}\alpha) + \frac{e^{\bar{\phi}}}{a}\bar{\mu}\dot{\bar{\phi}}[\dot{h} + 6\dot{\varphi} + 2k^2(1-e^{4\bar{\phi}})\alpha] \\
    & + 8\pi Gae^{-3\bar{\phi}}[\delta\rho + 3\delta P - 3(\bar{\rho} + 3\bar{P})\varphi].
\end{split}
\end{align}
The equations for the perturbed vector field obey
\begin{align}
\dot{\alpha} =& E - \varphi + \bigg(\dot{\bar{\phi}} - \frac{\dot{a}}{a}\bigg) \alpha, \\
K_B\bigg(\dot{E} + \frac{\dot{b}}{b}E\bigg) =& -\bar{\mu}\dot{\bar{\phi}} (\varphi - \dot{\bar{\phi}}\alpha) + 8\pi Ga^2(1-e^{-4\bar{\phi}}) (\bar{\rho} + \bar{P}) (\theta - \alpha).
\end{align}

The perturbed modified Einstein equations yield
\begin{align}
\begin{split}
& 2k^2(\varphi - \eta) + e^{4\bar{\phi}}\frac{\dot{b}}{b} \bigg(\dot{h} + 2k^2(1-e^{-4\bar{\phi}})\alpha + 6\frac{\dot{a}}{a}\varphi\bigg) + ae^{3\bar{\phi}} \bigg(\dot{\bar{\phi}} - \frac{3}{U}\frac{\dot{b}}{b}\bigg) \gamma \\
& - K_Bk^2E = 8\pi Ga^2\bar{\rho} (\delta - 2\varphi),
\end{split} \\
& 2k^2\dot{\eta} - 2k^2 \bigg(\frac{\dot{a}}{a} + \bar{\mu}\dot{\bar{\phi}}\bigg)\varphi + \frac{k^2}{U}ae^{-\bar{\phi}}\gamma = 8\pi Ga^2e^{-4\bar{\phi}} (\bar{\rho} + \bar{P}) k^2\theta. \label{eq.Einstein3}
\end{align}

To solve these perturbation equations, we need to specify
the initial conditions. In \cite{Skordis2008}
the adiabatic initial conditions of scalar mode
perturbations during the radiation era were proposed.
In our numerical computations in the following
discussions we adopt those initial
conditions for the selected special potential
(\ref{eq.VBek}).

\section{Large scale structure in TeVeS theory}

\subsection{The growth of the baryon density fluctuation}\label{sec.growth}

The growth of structure in TeVeS theory was first
discussed in \cite{Skordis2006a}. It was reported
that with the decrease of the TeVeS parameters $K_B$, $l_B$ and $\mu_0$,
the small scale power spectrum of the baryon
density fluctuations can be boosted to mimic that
in the adiabatic $\Lambda$CDM model. In
\cite{Dodelson2006} it was pointed out that the
growth of structure in the TeVeS is mainly due to the vector
field. In \cite{Skordis2009}, it was further clarified
that even if the contribution of the TeVeS fields
to total energy budget in the background FLRW universe is negligible,
we can still have a growing mode of density fluctuations that drives
structure formation.

Although the matter power specturm in TeVeS theory can mimic that in $\Lambda$CDM
cosmology, the mechanism of structure
growth in two models is different. In the
$\Lambda$CDM model, after decoupling from photons,
baryons fall into the gravitational wells induced
by CDM. In TeVeS theory, the growth of
perturbations is mainly driven by the vector
field that grows rapidly after recombination.
Thus it may be possible to distinguish them by
studying the evolution history of the perturbations.

\begin{figure*}[tb!]
\subfloat[]{
    \includegraphics[width=0.45\textwidth]{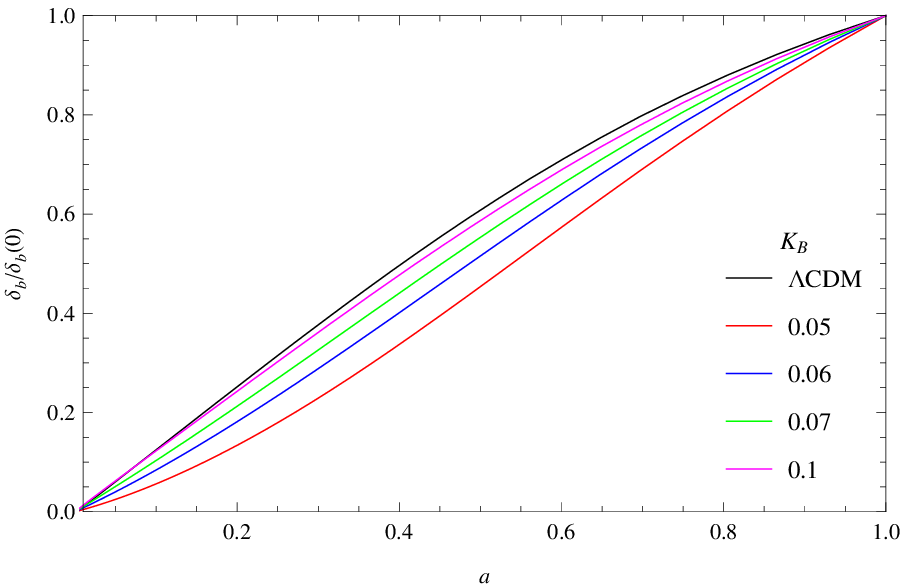}
}
\subfloat[]{
    \includegraphics[width=0.45\textwidth]{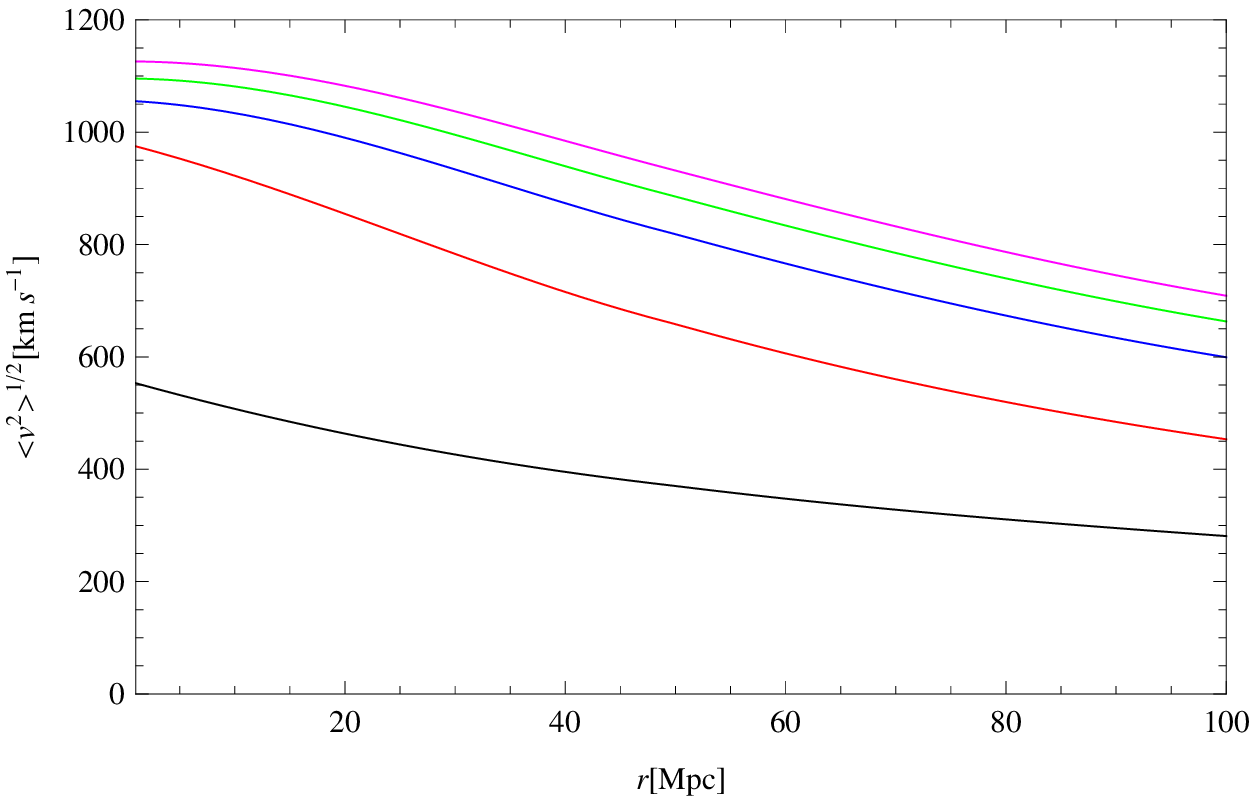}
}\\
\subfloat[]{
    \includegraphics[width=0.45\textwidth]{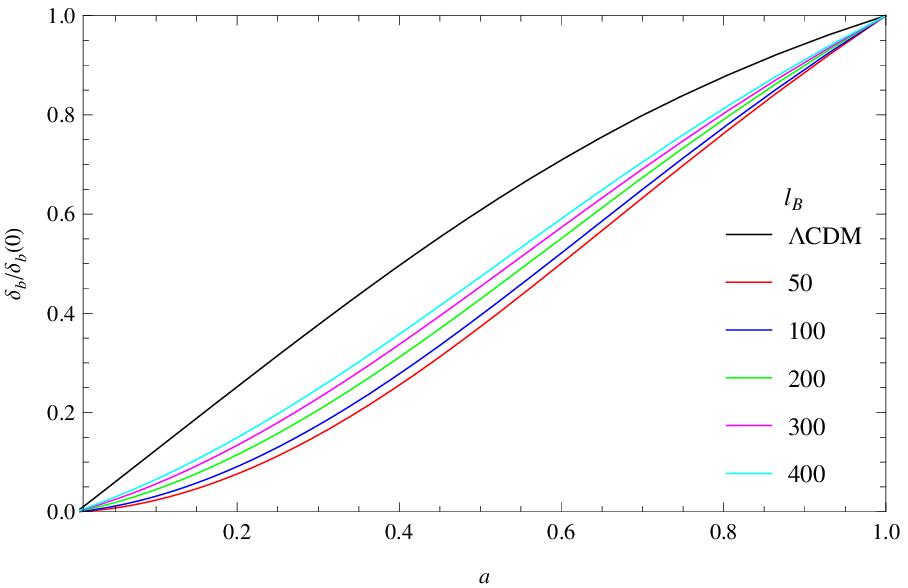}
}
\subfloat[]{
    \includegraphics[width=0.45\textwidth]{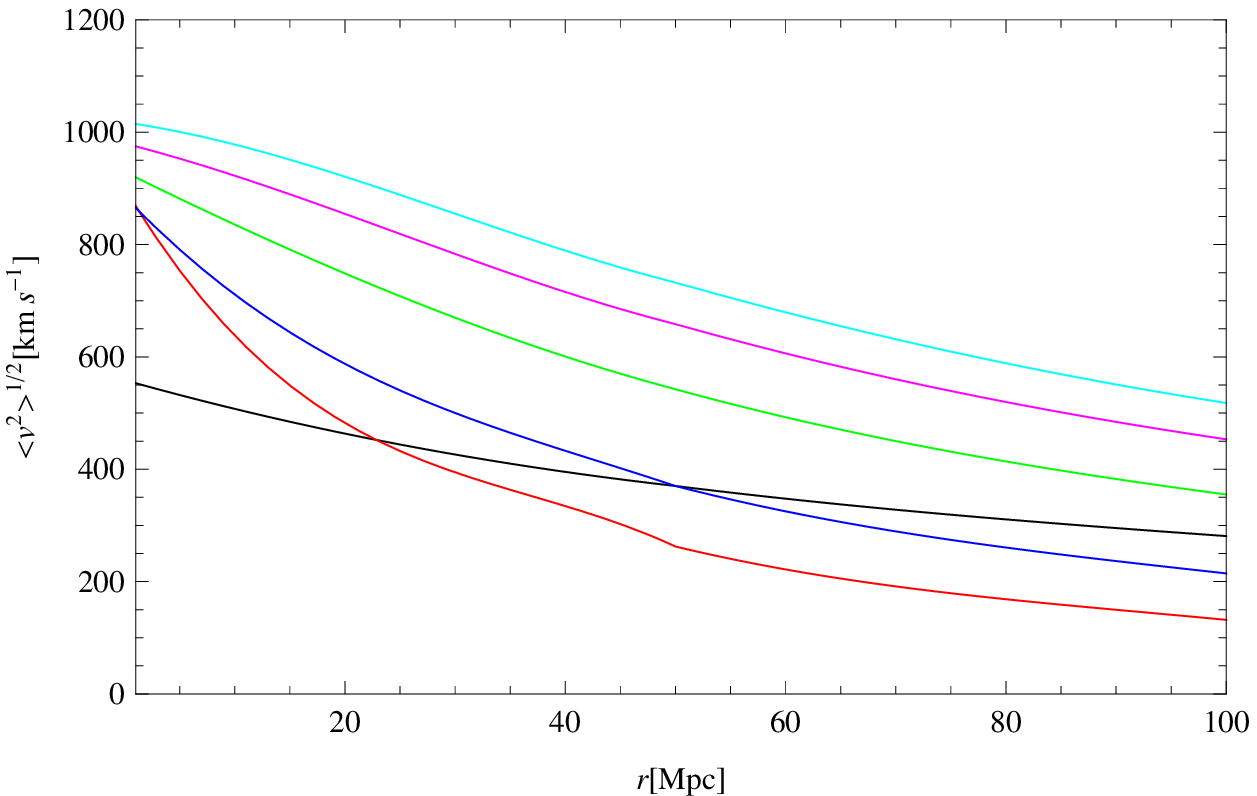}
}\\
\subfloat[]{
    \includegraphics[width=0.45\textwidth]{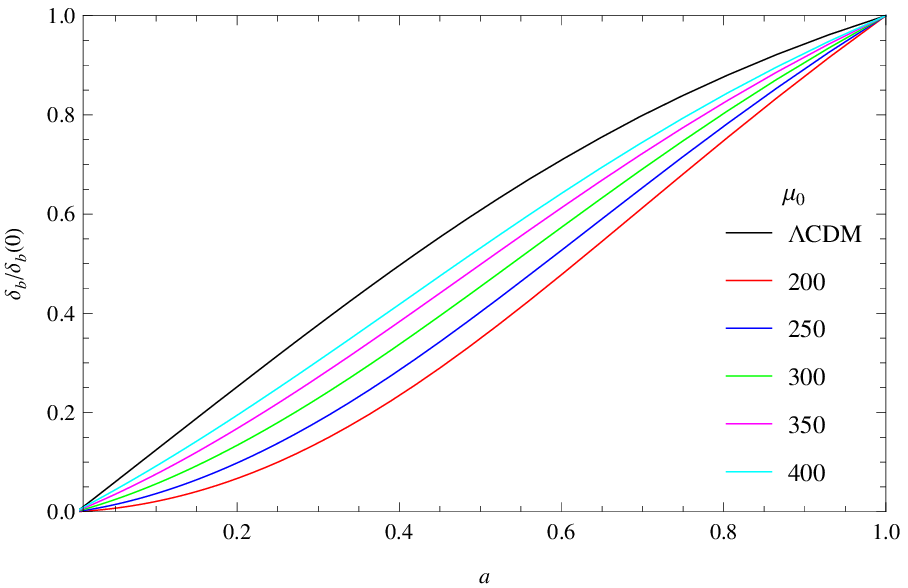}
}
\subfloat[]{
    \includegraphics[width=0.45\textwidth]{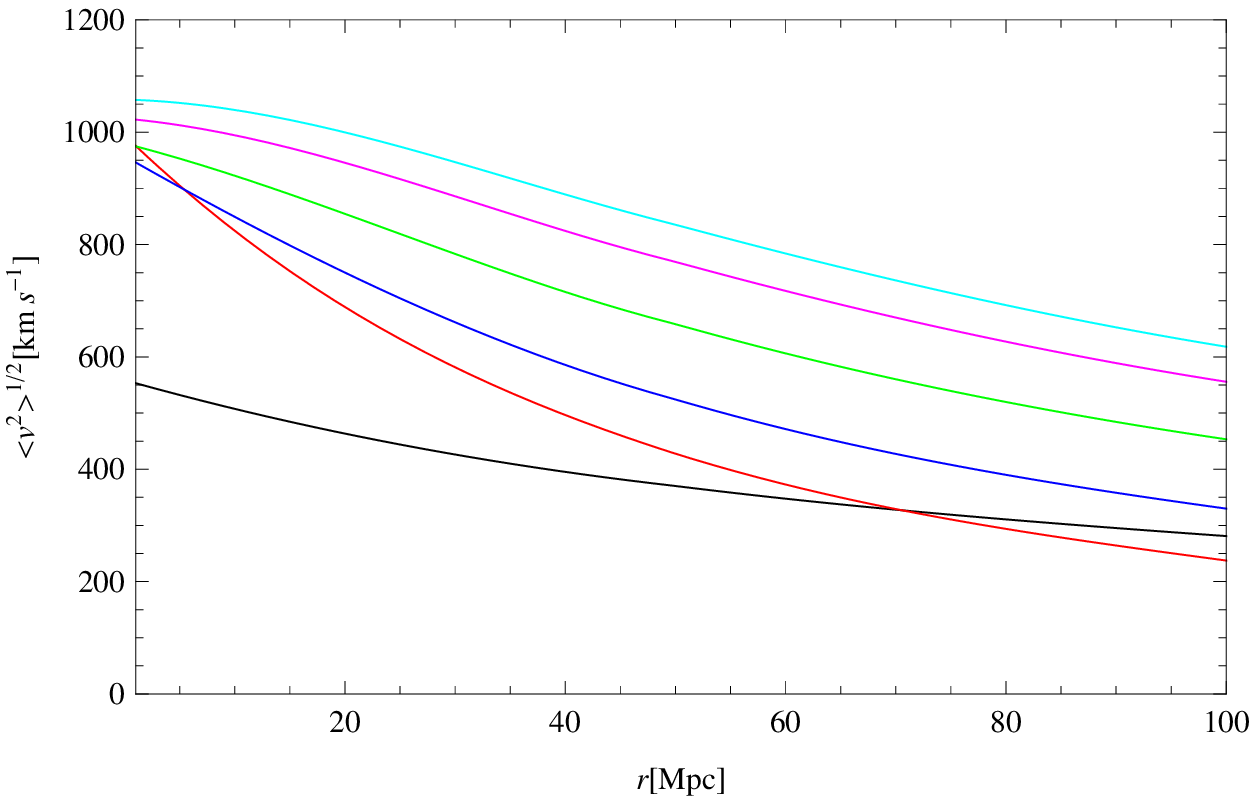}
}
\caption{\label{fig.delta} The figures on the left column are the
evolutions of $\delta_{\mathrm{b}}$ for
$k=0.1\mathrm{Mpc}^{-1}$, normalized to its
present value; on the right column is the root mean square(rms)
dispersion of baryon peculiar velocities. The
black curves correspond to the fiducial $\Lambda$CDM model.
The colored curves are for TeVeS models with
different parameters. The curves in the figures on the
same row follow the same convention. The TeVeS
parameters are $K_B=0.05$, $l_B=300$ and
$\mu_0=300$ if not specified. The fourth neutrino abundance is $\Omega_{\nu}h^2=0.15$}
\end{figure*}

In the left column of Fig.\ref{fig.delta}, we
demonstrate the evolutions of baryon density
perturbation in synchronous gauge, $\delta_{\mathrm{b}}$,
in TeVeS models. The density perturbations are evaluated for
$k=0.1\mathrm{Mpc}^{-1}$. For comparison,
we also plot the evolution of $\delta_{\mathrm{b}}$ in the fiducial
$\Lambda$CDM model. We take the
cosmological parameters $\Omega_{\mathrm{b}} h^2
= 0.022$, $\Omega_c h^2 = 0.12$, $h = 0.68$,
$\tau = 0.09$; $n_s = 0.96$ and
$\mathrm{ln}(10^{10}A_s) = 3.1$, where the Hubble
constant $H_0 = 100h
\mathrm{km}\cdot\mathrm{s}^{-1}\cdot\mathrm{Mpc}^{-1}$,
$\tau$ is the optical depth to the last
scattering surface, $n_s$ and $A_s$ are the
spectral index and amplitude of the primordial
power spectrum. We use these parameters for
the fiducial $\Lambda$CDM model and TeVeS models (except $\Omega_c h^2 = 0$) throughout this
paper.  We also introduce one sterile neutrino in addition to three massless neutrinos in the TeVeS cosmology, as \cite{Angus2008} suggested in order to fit the CMB observations. In the calculations $\Omega_{\nu} = 0.15$. As we expected, the growth rate of
$\delta_{\mathrm{b}}$ in TeVeS theory differs from
that in the $\Lambda$CDM model. In most cases, the
perturbations grow faster in TeVeS theory than in the $\Lambda$CDM model at low redshifts. And the
smaller the TeVeS parameters are, the more the growth rate deviates from the $\Lambda$CDM model. The growth rate is especially
sensitive to $K_B$ when it is small. Thus observing the structure growth
can also help in constraining the TeVeS
parameters.

Besides the evolution of the growth rate,
its spatial dependence is also attractive in distinguishing TeVeS cosmology from
$\Lambda$CDM. We discuss this topic in the
last subsection below.

\subsection{The peculiar velocity}\label{sec.velocity}

The peculiar velocity is related to the time
derivative of the density perturbation in the linear perturbation theory. In
Newtonian gauge, we have the relation
\begin{equation}
v_{\mathrm{b}}^{(N)} = -\frac{\dot{\delta}_{\mathrm{b}}^{(N)}}{k} = -aHf^{(N)}\frac{\delta_{\mathrm{b}}^{(N)}}{k},
\label{eq.v}
\end{equation}
where $f^{(N)} \equiv
\frac{\mathrm{dln}\delta_{\mathrm{b}}^{(N)}}{\mathrm{dln}a}$ is the linear
growth factor and `$N$' means that the quantity is evaluated in Newtonian gauge. For conciseness, we will omit `$N$' in $v_{\mathrm{b}}^{(N)}$ in the following.

To estimate the magnitude of $v_{\mathrm{b}}$, we
first solve Eqs. (\ref{eq.Ddelta})-(\ref{eq.Einstein3}) and derive
the peculiar velocity of baryon in Newtonian
gauge. Then we compute the root mean square (rms)
dispersion of $v_{\mathrm{b}}$ within a sphere of
radius $r$ by
\begin{equation}
\langle v_{\mathrm{b}}^2 \rangle = \int \mathrm{d}^3k W_r^2(k) P_v(k),
\end{equation}
where $W_r(k)$ is a top hat window function of
radius $r$ and $P_v(k)$ is the power spectrum of
$v_{\mathrm{b}}$. The
magnitude $\langle v_{\mathrm{b}}^2
\rangle^{1/2}$ represents the mean velocity of
baryons within a sphere of radius $r$ with
respect to the mean matter distribution. For
comparison, we also compute the same magnitude
for the fiducial $\Lambda$CDM model.

We present the calculated $\langle v_{\mathrm{b}}^2
\rangle^{1/2}$ at $z=0.1$ in the right column of
Fig.\ref{fig.delta}. We see that the velocity in the
TeVeS model is larger than that in
$\Lambda$CDM at the scale of 10Mpc, which is
consistent with the fast growth rate displayed in
the left column. Depending on the parameters,
$v_{\mathrm{b}}$ in TeVeS can be as large as twice the $\Lambda$CDM value. With the
increase of radius $r$, the velocity dispersion in the TeVeS
model decays faster than in the $\Lambda$CDM model. When $r$ reaches 100 Mpc, the
velocity of the TeVeS models with small values of
$l_B$ and $\mu_0$ can become lower than that of
$\Lambda$CDM. In general smaller TeVeS
parameters lead to lower velocity. This may be counterintuitive since $\delta_{\mathrm{b}}^{(N)} \simeq \delta_{\mathrm{b}}$ grows faster for smaller TeVeS parameters. This is
because $v_{\mathrm{b}}$ is proportional to the density
perturbation $\delta_{\mathrm{b}}^{(N)}$ as well as the growth factor. With the decrease of TeVeS parameters, $\delta_{\mathrm{b}}^{(N)}$ is getting smaller. At
$z=0.1$, the influence of low density fluctuation overwhelms the high growth rate of baryons and the net effect is the decrease of $v_{\mathrm{b}}$.

Observationally it is difficult to measure the
peculiar velocity on scales above $50h^{-1}$Mpc using galaxies. The kSZ effect provides an alternative
method of great promise to measure peculiar velocity at cosmological
distances,  without resorting to distance indicators. High
resolution and low noise CMB experiments have the potential to
measure various statistical averages of cluster velocity such as the
bulk flow (e.g. \cite{Kashlinsky2000, Atrio-Barandela2004}), the mean pairwise momentum
(e.g. \cite{Hand2012}) and the momentum power spectrum (e.g. \cite{Zhang2008a}). Advanced CMB experiments even have the capability of
measuring the peculiar velocity of individual galaxy clusters (e.g. \cite{Haehnelt1996, Aghanim2001, Holder2004, Knox2004, Aghanim2005, Diaferio2005}). In \cite{Hand2012} Hand {\it et al.} reported the measurement of the mean pairwise momentum of clusters using the CMB sky
map made by the Atacama Cosmology Telescope(ACT).
{\it Planck} found the radial peculiar velocity
rms to be below three times the $\Lambda$CDM
prediction at $z=0.15$ \cite{PlanckCollaboration2013a}.
While the results from ACT and {\it Planck} seem
to be consistent with the $\Lambda$CDM model,
given their large uncertainties they are also
compatible with TeVeS cosmology. To conclude,
while at present the data do not have the
statistical power to constrain the TeVeS
parameters, the peculiar velocity field could
become an important test of TeVeS theory with
future data sets of higher resolution and lower
noise.

\subsection{The kinetic Sunyaev-Zel'dovich effect}\label{sec.ksz}

The Sunyaev-Zel'dovich(SZ)
effect\cite{Sunyaev1972} is generated through the
scattering of CMB photons by free electrons while
the photons travel through ionized gas after
reionization. The SZ effect is commonly
classified into two sorts: the thermal SZ (tSZ)
effect, which is characterized by the thermal
motion of free electrons, and the kSZ effect, which is
characterized by their bulk motion. Because free electrons produced after
reionization of the intergalactic medium
share the same motion as the plasma, it
is expected that the kSZ effect can serve as a
probe of baryon peculiar velocity field.

The kSZ effect induces distortions on the CMB
temperature map. The kSZ temperature anisotropy is given by
\begin{equation}
\frac{\Delta T(\hat{\bm n})}{T_{\mathrm{CMB}}} =
-\int_{t_{\mathrm{re}}}^{t_0} n_{\mathrm{e}}\sigma_Te^{-\kappa}(\bm{v}_{\mathrm{e}}^{(N)}\cdot\hat{\bm n})\mathrm{d}t,
\end{equation}
where $n_{\mathrm{e}}$ is the electron density,
$\sigma_T$ is the Thomson cross section and
$\kappa$ is the Thomson optical depth, and $\bm{v}_{\mathrm{e}}^{(N)}$
is the peculiar velocity of free electrons; the
integral is along the line of sight (l.o.s.) out
to the reionization epoch and $\hat{\bm n}$ is
the unit vector along the l.o.s. The contribution
of the kSZ effect to the CMB temperature angular
power spectrum is\cite{Vishniac1987, Ma2002, Zhang2004}
\begin{equation}
C_l^{kSZ}=\frac{16\pi^2}{(2l+1)^3} (\bar{n}_{\mathrm{e}}(0)\sigma_T)^2
\int^{z_{\mathrm{re}}}_0 (1+z)^4 \chi^2_{\mathrm{e}} \frac{1}{2}\Delta^2_{\mathrm{B}}(k,z)\rvert_{k=l/x}
e^{-2\kappa} x(z) \frac{\mathrm{d}x(z)}{\mathrm{d}z}\mathrm{d}z,
\end{equation}
where $x$ is the comoving distance, $\bar{n}_{\mathrm{e}}(0)$ is
the mean electron number density at present,
$\chi_{\mathrm{e}}$ is the ionization fraction
and $\Delta_{\mathrm{B}}^2(k,z) \equiv
\frac{k^3}{2\pi^2}P_{\mathrm{B}}(k,z)$.
$P_{\mathrm{B}}$ is the power spectrum of the
curl part of $p \equiv (1+\delta_{\mathrm{e}}^{(N)})\bm{v}_{\mathrm{e}}^{(N)}$. In the
linear regime, $\bm{v}_{\mathrm{e}}^{(N)}$ is curl free and only the
combination $\delta_{\mathrm{e}}^{(N)}\bm{v}_{\mathrm{e}}^{(N)}$ contributes to
$P_{\mathrm{B}}$. Given $\delta_{\mathrm{e}}^{(N)}=\delta_{\mathrm{b}}^{(N)}$, $\bm{v}_{\mathrm{e}}^{(N)}=\bm{v}_{\mathrm{b}}$ and (\ref{eq.v}), $P_{\mathrm{B}}$ can be
written as
\begin{equation}\begin{split}
P_{\mathrm{B}}(k,z) =
& \frac{1}{2} \int \frac{\mathrm{d}^3\bm{k}'}{(2\pi)^3}
\bigg(\frac{\dot{D}(z)}{D(z)}\bigg)^2 P(k',z) P(k-k',z)\\
&\times [W_g(k-k')\beta(\bm{k},\bm{k}') + W_g(k')\beta(\bm{k},\bm{k}-\bm{k}')]^2,
\end{split}
\label{eq.P_B}
\end{equation}
where $D(z) \equiv \delta_{\mathrm{b}}^{(N)}(z) /
\delta_{\mathrm{b}}^{(N)}(0)$ is the growth function of the
baryon, $P(k)$ is the baryon power spectrum in Newtonian gauge,
$W_g(k)$ is the transfer function that takes
into account the suppression of baryon density
fluctuations at small scales due to physical
processes\cite{Fang1993}, and
$\beta(\bm{k},\bm{k}') =
[\bm{k}'-\bm{k}(\bm{k}\cdot\bm{k}')/\bm{k}^2]/\bm{k}'^2$.
For simplicity, we have set $W_g(k)$ to unity in
our numerical calculations.

The nonlinear evolution of density perturbations
enhances the power spectrum at small scales. To
account for this effect, we rewrite
(\ref{eq.P_B}) into \cite{Hu2000, Ma2002, Shaw2011}
\begin{equation}
\begin{split}
P_{\mathrm{B}}(k,z) =
& \frac{1}{2} \int \frac{\mathrm{d}^3\bm{k}'}{(2\pi)^3}
\bigg(\frac{\dot{D}}{D}\bigg)^2 P(k',z) P(k-k',z)\\
&\times [W_g(k-k')T_{NL}(k-k')\beta(\bm{k},\bm{k}')
    + W_g(k')T_{NL}(k')\beta(\bm{k},\bm{k}-\bm{k}')]^2,
\end{split}
\label{P_B_NL}
\end{equation}
where we have defined the nonlinear power
spectrum as $P^{NL}(k) \equiv P(k)T_{NL}^2(k)$.
It is assumed that the nonlinear corrections
affect the density perturbation only and the
velocity field is still linear\cite{Coles1991}. \cite{Zhang2004} found that the
other linear power spectrum should also be
replaced by the nonlinear one to better describe the simulated
$\Delta^2_B$. This is likely caused by the extra contribution from the
curl velocity component generated by shell crossing. To include the
nonlinear correction we need to specify $T_{NL}(k)$ for the TeVeS model, which is usually done by using
adequate fits to N-body simulations. However,
such a simulation has not been carried out in TeVeS
theory. It is then difficult to give a reliable
description of the nonlinear corrections. As a
first guess, we borrow the halofit fitting
formula\cite{Smith2003, Takahashi2012} for
$\Lambda$CDM model to evaluate the nonlinear
power spectrum. We have to emphasize that this is
only a rough estimation because TeVeS theory
is significantly different from GR at cluster
scales where the kSZ effect becomes important in
the CMB anisotropies.

\begin{figure*}[tp]
\subfloat[]{
    \includegraphics[width=0.45\textwidth]{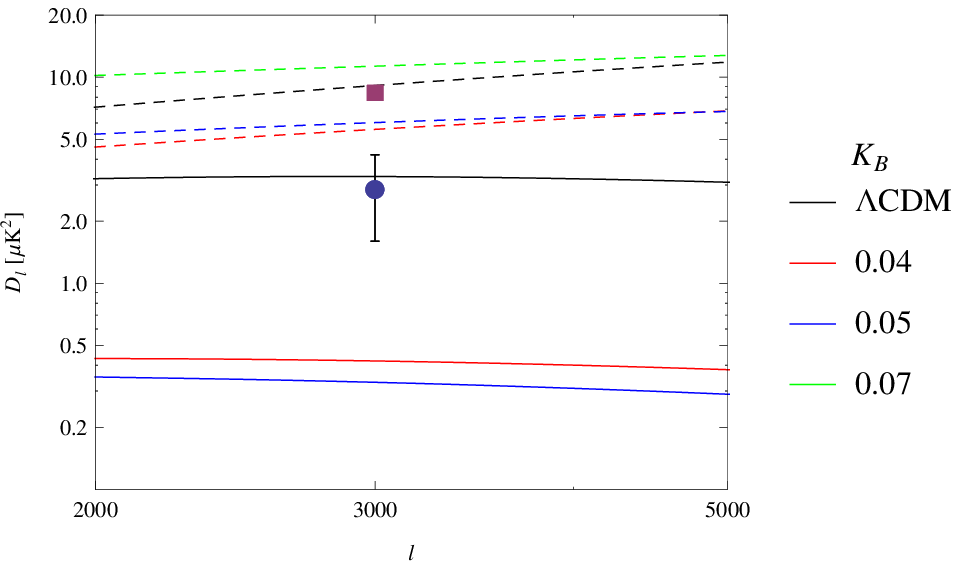}
}
\subfloat[]{
    \includegraphics[width=0.45\textwidth]{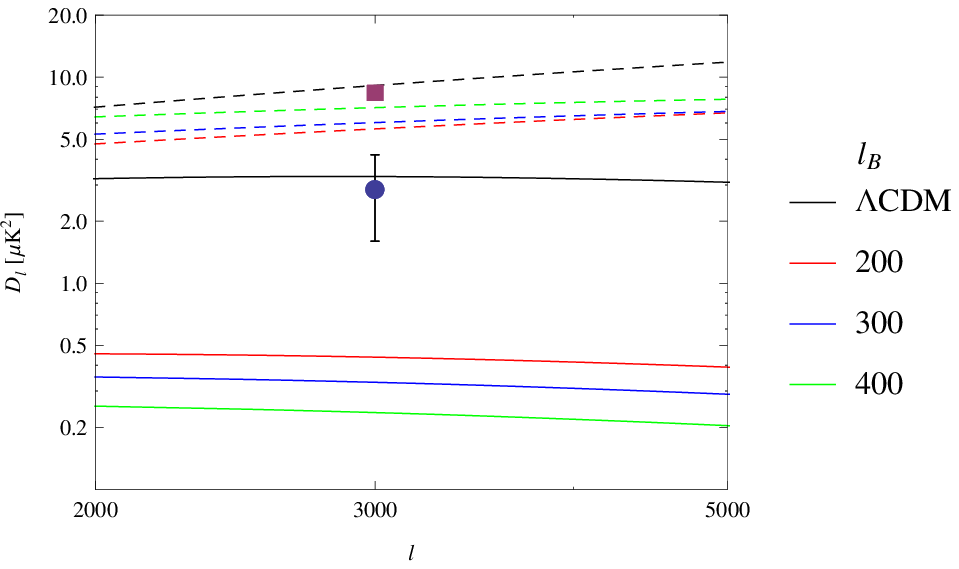}
}\\
\subfloat[]{
    \includegraphics[width=0.45\textwidth]{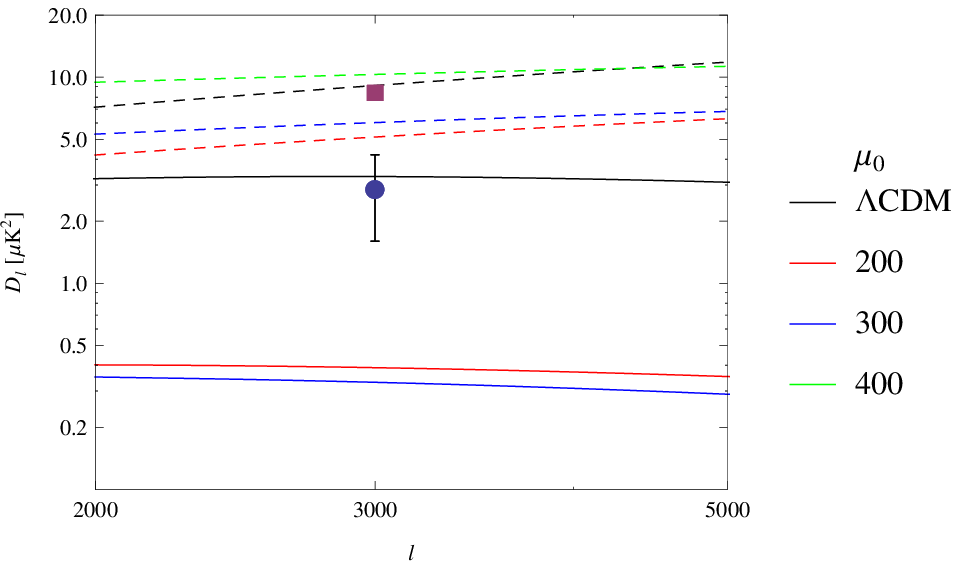}
} \caption{\label{fig.ksz} The kSZ anisotropy power spectra for the
TeVeS model with different parameters. The black
curve is for the fiducial $\Lambda$CDM model. The
solid lines represent the linear kSZ power
spectra and the dashed lines are for the kSZ
power spectra taking into account the nonlinear
corrections. The TeVeS parameters are the same as in Fig.\ref{fig.delta} if not
specified. }
\end{figure*}

In Fig.\ref{fig.ksz}, we present the theoretical predictions
of both linear and nonlinear kSZ power spectrum
in the TeVeS model and the fiducial $\Lambda$CDM
model. For consistency, we have assumed $\tau=0.09$ for all models. The solid lines represent the linear kSZ
effect. The power spectra for TeVeS are always smaller
than that of the $\Lambda$CDM model. Increasing
the TeVeS parameters will further suppress the
kSZ effect. Taking into account the nonlinear
effect, the power spectra for TeVeS are enhanced and
become comparable with the $\Lambda$CDM
model. In contrast to the linear kSZ effect,
increasing the TeVeS parameters enhances the power
spectrum. The difference may be the consequence of the
scale-dependent evolution of perturbations in
TeVeS. $T_{NL}(k)$ varies with $k$, which means
that the main contributions to the linear and
nonlinear kSZ power spectrum come from different
scales. And the linear matter power spectrum
$P(k)$ at different scales changes differently
when the parameters vary. Therefore the linear
and nonlinear power spectra respond differently to the
changing of the parameters. Again we emphasize that
this phenomenon depends heavily on the estimation
of $P^{NL}(k)$, and it is premature to make solid
conclusion before we can have an accurate nonlinear
matter power spectrum in TeVeS theory.

We include two data points for the kSZ power spectrum
in Fig.\ref{fig.ksz}. The rectangle indicates the
upper limit of $D_l \equiv l(l+1)C_l/2\pi$ at $l=3000$ with 95\% C.L.
derived from ACT data, $D_{3000}^{\mathrm{kSZ}} <
8.6 \mu\mathrm{K}^2$\cite{Sievers2013}. The
circle with the error bar indicates the measurement
of the SPT-SZ survey using data from the South
Pole Telescope(SPT), $D_{3000}^{\mathrm{kSZ}} =
2.9 \pm 1.3 \mu\mathrm{K}^2$ with 68\% C.L.\cite{George2014}.
These measurements heavily rely on modeling of cosmic infrared
background and tSZ contributions, and therefore suffer from
significant systematic uncertainty.
Meanwhile our theoretical predictions have
considerable uncertainties. Besides the
nonlinear effect, we have assumed a simple
instantaneous reionization model with $\tau=0.09$
while the kSZ effect from the patchy reionization
is expected to be important. Hence the kSZ power spectrum may be underestimated.  Besides, that $\tau$ in all models have the same value is a rough assumption itself, since a change in the rate of structure growth will also change the optical depth. On the
other hand, we did not include the smoothing in the
gas density caused by the gas pressure in our calculation, which could
potentially reduce the amplitude of the kSZ power spectrum.  And it is known that some fraction of the electrons is locked up in stars and neutral clouds, which further reduces the kSZ amplitude. Despite these uncertainties, our
computations indicate that the linear kSZ power
spectrum in the TeVeS model is consistent with the upper limits of the
observations. The fact that it is smaller than the lower limit of SPT measurement does not rule out the TeVeS since the linear kSZ power spectrum is essentially a lower limit to the realistic one. But if we look at the nonlinear kSZ
power spectra, they are certainly ruled out by the SPT
observation, and the ACT measurement puts a tight constraint on the model parameters.

\subsection{The scale dependence of growth rate}\label{sec.scaleDepend}

One of the characteristic features of the $\Lambda$CDM model is the
scale-independent growth rate in the subhorizon
approximation\cite{Zhang2011}. This property was
found to be violated if the gravity goes beyond
GR\cite{Brans1961, Dvali2000, Nicolis2009,
Tsujikawa2011, Clifton2012}, if DE clustering
cannot be neglected\cite{Creminelli2009, Parfrey2011} or if DE couples to DM\cite{He2009a, He2010}. Now we investigate
the scale dependence of the growth rate in
TeVeS theory and see whether it can serve to
distinguish the TeVeS from the $\Lambda$CDM model.

\begin{figure*}[tp]
\subfloat[]{
    \includegraphics[width=0.45\textwidth]{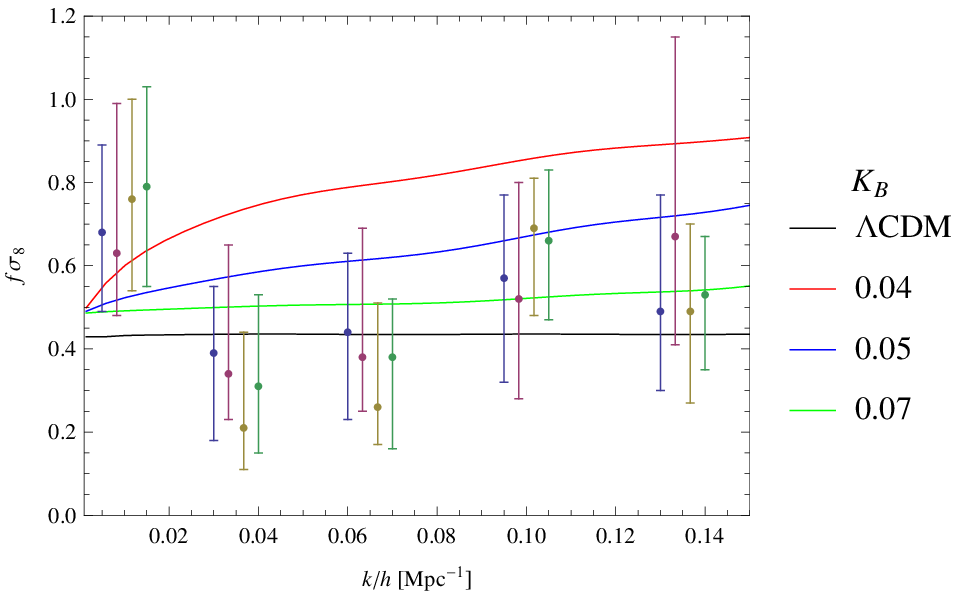}
}
\subfloat[]{
    \includegraphics[width=0.45\textwidth]{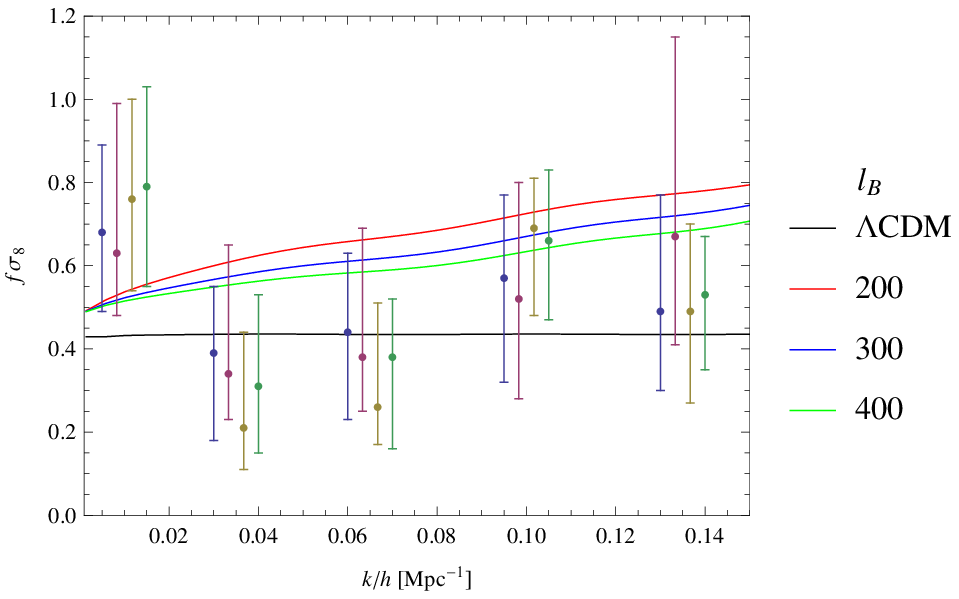}
}\\
\subfloat[]{
    \includegraphics[width=0.45\textwidth]{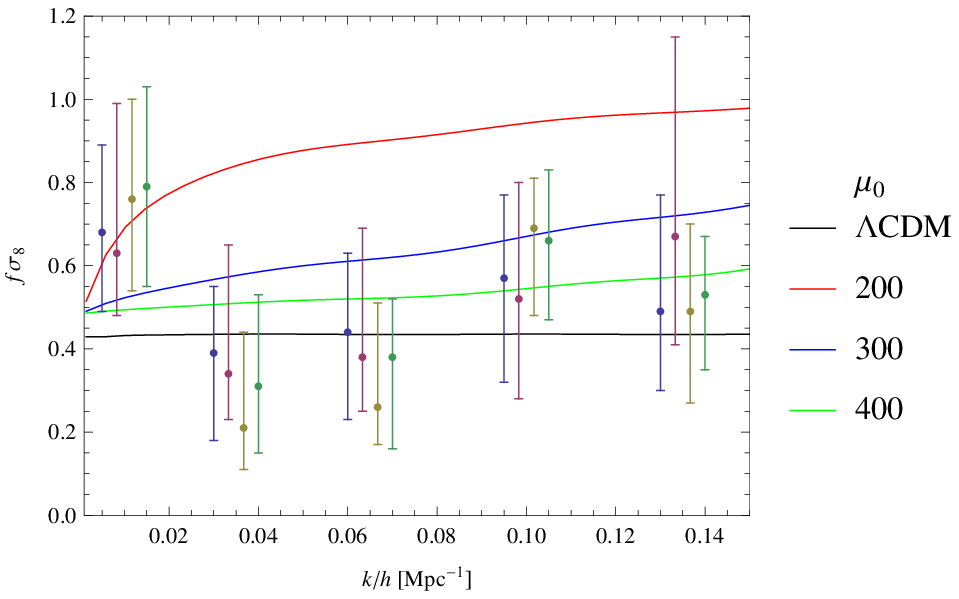}
}
\caption{\label{fig.fs8} $f\sigma_8(k;z=0)$ for the TeVeS model
with different parameters. The black curve is for
the fiducial $\Lambda$CDM model. The data points
are from the 6-degree Field (6dF) Galaxy Survey. For all curves,
$\sigma_8(0) = 0.834$.}
\end{figure*}

Since observations are in fact sensitive to the normalized growth rate $f\sigma_8(k;z)$ instead of $f(k;z)$, in Fig.\ref{fig.fs8} we display $f\sigma_8(k)$ in synchronous gauge for baryon with respect to $k/h$ at redshift $z=0$. In all models, we use the same value for $\sigma_8(0)$, $\sigma_8(0) = 0.834$; thus, we can concentrate on the scale dependence of the growth rate. It is a realistic assumption, since $\sigma_8(0)$ in a viable cosmology model should be similar to that in the fiducial $\Lambda$CDM model. The
growth rate in TeVeS is systematically higher
than that in the $\Lambda$CDM model. The black
curve for the fiducial $\Lambda$CDM model is
almost a horizontal line, reflecting the
scale-independent growth of density
perturbations. In contrast to the $\Lambda$CDM
model, $f\sigma_8(k)$ in TeVeS theory clearly varies with scale. We see that the growth
rate is bigger at small scales than large scales. Increasing the TeVeS
parameters, $f\sigma_8(k)$ at given $k$ becomes smaller, which
is consistent with the behavior seen in
Fig.\ref{fig.delta}. Furthermore, $f\sigma_8(k)$ converges for different parameters
when $k \rightarrow 0$, if $\sigma_8$ is
equally normalized.

We compare the theoretical prediction of
$f\sigma_8(k;z=0)$ with the measurement using the
observations of peculiar motions of galaxies of the
6-degree Field(6dF) Galaxy Survey velocity sample
together with a newly compiled sample of
low-redshift type Ia
supernovae\cite{Johnson2014}. The measurement
was done in 5 $k$ bins: $k_1=[0.005,0.02]$,
$k_2=[0.02,0.05]$, $k_3=[0.05,0.08]$,
$k_4=[0.08,0.12]$ and $k_5=[0.12,0.15]$. The data
points in different color refer to results
derived by different data sets and
methodologies. The measurement does not show
strong evidence for a scale dependence in the
growth rate. But we see that the TeVeS prediction
matches the measured $f\sigma_8$ for a wide range
of parameters.

Currently, the measurements of the scale dependence of growth rate is not as accurate as the average growth rate at different redshifts measured through redshift-space distortion observations. The latter has been used in the literature to constrain cosmological models(e.g. \cite{Tsujikawa2012,Macaulay2013,Salvatelli2014}). In this paper, we concentrated on the scale dependence of growth rate and hope that future precise data  can help to constrain the TeVeS  cosmological model.

\section{Cosmic microwave background radiation in TeVeS theory}\label{sec.cmb}

In the last section, we investigated the structure growth in TeVeS cosmology. On the baryon peculiar velocity, kSZ effect, and scale dependence of growth rate, the theoretical predictions all exhibit clear difference between tge TeVeS model and $\Lambda$CDM model. However in observations, current data are not precise and powerful enough to distinguish clearly the TeVeS model from the $\Lambda$CDM.
In this section, we turn to study the CMB power spectrum in TeVeS cosmology. CMB experiments probe larger scales and deeper redshift of the Universe than large scale structure observations. Meanwhile, precise measurements of the CMB have been available. They can be used to tightly constrain the TeVeS model.

In \cite{Skordis2006a}, the first numerical
calculation of CMB angular power spectrum in
TeVeS theory was done by using the original
Bekenstein's potential (\ref{eq.VBek}). The authors
found that a flat universe composed of about 5\%
baryon and 95\% cosmological constant today
matches the observations poorly. The angular
distance relation was found modified as compared to the
standard adiabatic $\Lambda$CDM universe. The
positions of the peaks in the CMB angular power
spectrum were observed  shifted to higher $l$s which led to a severe mismatch with the
observational data. This problem was argued to be
cured if the three neutrinos have a mass of $m_{\nu} \simeq 2\mathrm{eV}$\cite{Skordis2006a}. In
\cite{Angus2008} it was argued that if including a
sterile neutrino with $\Omega_{\nu} \simeq 0.23$ ($m_{\nu} \simeq 11\mathrm{eV}$) in addition to the three massless neutrinos,
the peaks of the CMB power spectrum will be located at
the right positions to match the observational
data. Furthermore by fitting a MOND-like
model to the WMAP five year data, it was concluded
that the  model with the sterile neutrino is
compatible with the observation. But in \cite{Angus2008}, it was assumed that there were no MOND
effects before recombination; therefore, the MOND effects have no influence on the CMB power spectrum. It
was commented that the fitting result in
\cite{Angus2008} has nothing to do with TeVeS
features\cite{Skordis2009}.

\begin{figure*}[tb]
\includegraphics[]{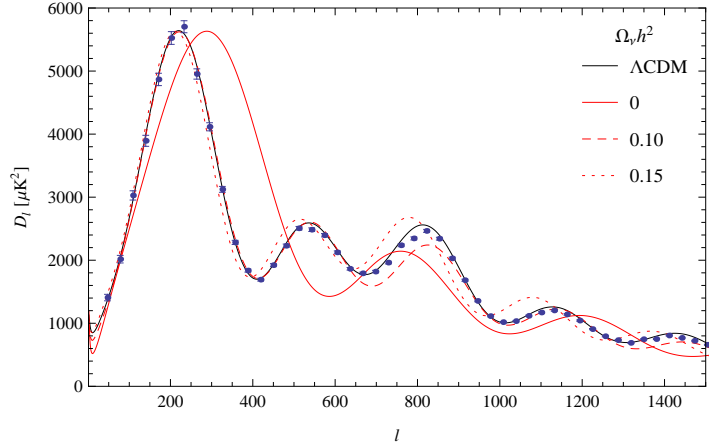}
\caption{\label{fig.cmb_mnu} The CMB temperature angular power
spectra for the fiducial $\Lambda$CDM
model(solid black curve) and TeVeS
models(red curves) having various amounts of
sterile neutrino. The data points with error bars
are from {\it Planck} 2013 results.}
\end{figure*}

Here, we do the whole calculation in the framework of TeVeS theory. We
numerically calculate the CMB power spectrum in
TeVeS theory in the presence of a sterile
neutrino. Our results are demonstrated in
Fig.\ref{fig.cmb_mnu}. The black line is for the
fiducial $\Lambda$CDM model. The red lines are for TeVeS models with
various $\Omega_{\nu} h^2$. To illustrate the
qualitative influence of the abundance of the
sterile neutrino, we fix the parameters in the
TeVeS models by taking $K_B = 0.1$, $l_B = 100$
and $\mu_0 = 300$. The other parameters are the
same as in the fiducial $\Lambda$CDM model except that we
have no CDM in TeVeS and $\mathrm{ln}(10^{10}A_s)$ is adjusted such that the first peaks of the power spectra have the same height. The data points and error bars
are from the {\it Planck} 2013
results\cite{PlanckCollaboration2013f}. It is
clear in Fig.\ref{fig.cmb_mnu} that including the
fourth neutrino can move the locations of the
acoustic peaks to larger angular scales.
Moreover, it can also enhance the third acoustic
peak to almost as high as the second peak, which
is usually considered the signature of CDM in the
Universe. With the increase of the abundance of
the fourth neutrino, there is clearly a
competition between the shift of the peak
positions and the enhancement of the third peak.

\begin{figure*}[tb]
\subfloat[]{
    \includegraphics[width=0.45\textwidth]{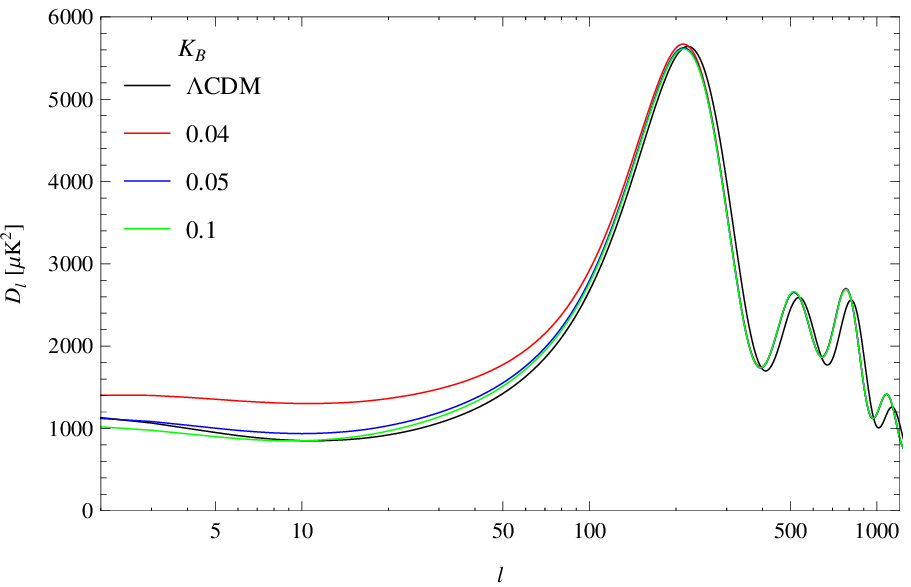}
}
\subfloat[]{
    \includegraphics[width=0.45\textwidth]{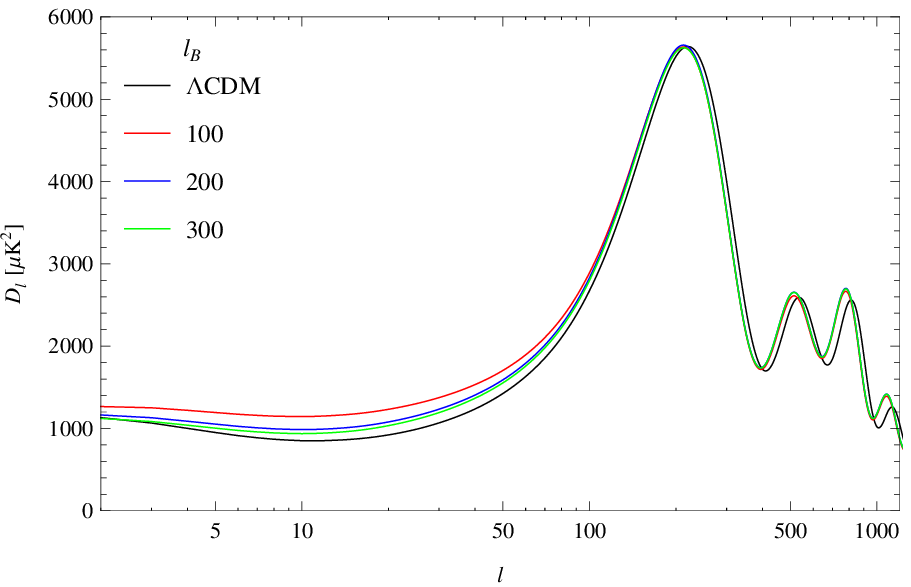}
}\\
\subfloat[]{
    \includegraphics[width=0.45\textwidth]{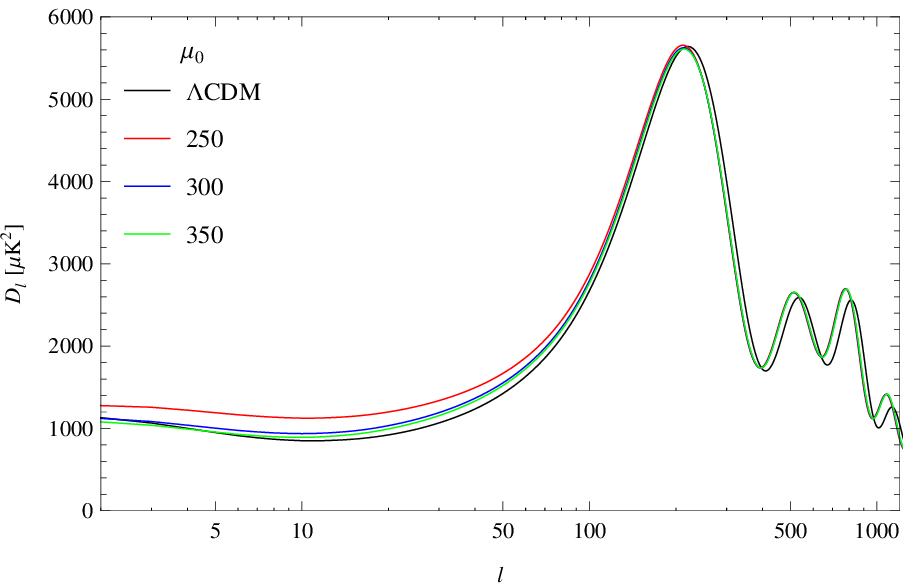}
}
\caption{\label{fig.cmb} The CMB temperature angular power
spectra for TeVeS models with different
parameters. The black curve is for the fiducial
$\Lambda$CDM model. The TeVeS parameters are
$K_B=0.05$, $l_B=300$ and $\mu_0=300$ if not
specified. The fourth neutrino abundance is $\Omega_{\nu}h^2=0.15$}
\end{figure*}

In \cite{Skordis2006a}, the authors found that changing
the TeVeS parameters will modify the CMB power
spectrum. It was observed that sufficiently small
TeVeS parameters, $K_B, l_B$ and $\mu_0$, can
cause the excess of the CMB power at large
scales. Their conclusion was obtained in the
absence of the sterile neutrino. We can see a
similar property in Fig.\ref{fig.cmb} where the
fourth neutrino has an abundance of $\Omega_{\nu}
h^2 = 0.15$. Smaller TeVeS parameters
consistently enhance the large scale power in
CMB. The CMB power spectrum at small $l$s is more
sensitive to the parameter $K_B$ than the other
two parameters. Considering that $K_B$ regulates
the dynamics of the vector field, our observation
here supports the argument in \cite{Dodelson2006}
that the vector field perturbation plays an
important role in the growth of structure in the TeVeS. Furthermore, we display in
Fig.\ref{fig.cmb} that the influence of TeVeS
parameters on the CMB power spectrum at small
scales is totally overshadowed by that of the abundance
of the fourth neutrino. The change of the
positions and amplitudes of acoustic peaks is
mainly caused by the change of $\Omega_{\nu}
h^2$.

\begin{table*}[tb!]
\caption{\label{tab.fit} The priors and fitting results of the cosmological parameters.}
\begin{tabular}{llll}
    \toprule
    Parameter ~~~~~~ & Best fit ~~~~~~ & 68\% limits ~~~~~~~~ & Prior \\
    \hline
    $K_B$                     & 0.0535 & $< 0.0701$ & [0.05, 0.5] \\
    $l_B$                     & 278 & $> 229$ & [10, 300] \\
    $\mu_0$                   & 326 & $329_{-41}^{+37}$ & [10, 400] \\
    $\Omega_{\nu} h^2$        & 0.157 & $0.156_{-0.002}^{+0.003}$ & [0.01, 0.5] \\
    $\Omega_{\mathrm{b}} h^2$            & 0.0209 & $0.0209 \pm 0.0002$ & [0.01, 0.03] \\
    $h$                       & 0.504 & $< 0.508$ & [0.5, 0.85] \\
    $\tau$                    & 0.00390 & $< 0.031$ & [0, 0.3] \\
    $n_s$                     & 0.898 & $0.900_{-0.007}^{+0.005}$ & [0.8, 1.4] \\
    $\mathrm{ln}(10^{10}A_s)$ & 2.89 & $2.93_{-0.06}^{+0.02}$ & [2.3, 3,5] \\
    \lasthline
\end{tabular}
\end{table*}

In order to test the viability of TeVeS theory in
explaining the observed CMB power spectrum and
constrain the TeVeS parameters as well as the
amount of the sterile neutrino, we confront the TeVeS
model with the {\it Planck} 2013
results\cite{PlanckCollaboration2013f}.  The data set we used includes the CMB TT power spectrum for $2 \leq l \leq 2500$. We
perform the numerical fitting using the Markov
chain Monte Carlo method. In the fitting,
we allow nine parameters to vary, which are $K_B$,
$l_B$, $\mu_0$, $\Omega_{\nu}h^2$, $\Omega_bh^2$,
$h$, $\tau$, $n_s$, and
$\mathrm{ln}(10^{10}A_s)$. The priors of these
parameters are listed in the last column in
Table \ref{tab.fit}. We modify the public code
CMBEASY\cite{Doran2005} to compute the CMB power
spectra and generate the Markov chains.

The TeVeS parameters have a similar influence on the
CMB. The CMB power spectrum for large $l$s hardly
depends on $K_B$, $l_B$ or $\mu_0$, while the
low-$l$ power is suppressed when one of the TeVeS
parameters increases. So one expects degeneracy
among them when fitting to the CMB observations.
Nevertheless, we can get moderate constraints for
them by using {\it Planck} data alone, as
indicated in Table \ref{tab.fit}.

Comparing with the $\Lambda$CDM model, the
constrained optical depth to the last scattering
surface in the TeVeS model is significantly smaller.
Inferring from {\it Planck} data, the 68\% C.L. for the optical depth is $\tau = 0.09 \pm
0.038$ for the $\Lambda$CDM
model\cite{PlanckCollaboration2013b}. Assuming
instantaneous reionization, the best-fit value
for the TeVeS model, $\tau = 0.0039$, implies that
reionization completed at $z = 1.2$. This is
certainly ruled out by astronomical observations
which suggests the end of reionization was at $z
\simeq 6$ or earlier. But if we take the 68\%
C.L. $\tau=0.31$, the situation becomes
better. The end of reionization was at $z=6.2$.
Thus, TeVeS cosmology is still marginally allowed
by current constraints of reionization history.

\begin{figure*}[tb]
\includegraphics[width=0.45\textwidth]{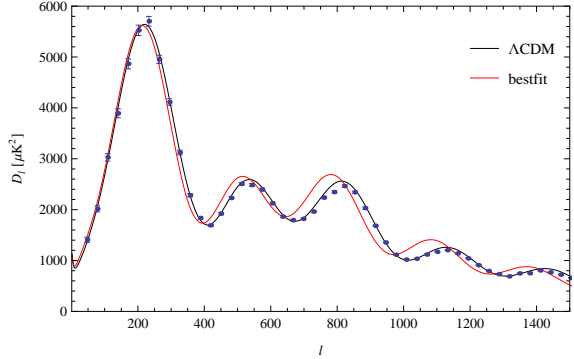}
\caption{\label{fig.cmb_best} The CMB temperature angular power
spectra for the best-fit TeVeS model (red) and the fiducial
$\Lambda$CDM model (black). The data points with error bars
are from {\it Planck} 2013 results. }
\end{figure*}

It is interesting that the constrained abundance of the sterile neutrino,
$\Omega_{\nu}h^2 = 0.156_{-0.002}^{+0.003}$, which corresponds to $m_{\nu} \simeq 15\mathrm{eV}$, is larger than the CDM
abundance($\simeq 0.12$) gotten in concordance with the
$\Lambda$CDM model\cite{PlanckCollaboration2013b}. Meanwhile, the obtained $h$ is much
smaller than the prediction in \cite{Angus2008}.
Recall that the locations of the acoustic peaks shift
towards smaller $l$ when $\Omega_{\nu}h^2$ increases; see Fig.\ref{fig.cmb_mnu}.
This explains why our constraint on $H_0$ is so
small. When Hubble's constant decreases, the
angular diameter distance to the last scattering
surface is increased. Thus, the angular scales of
the acoustic peaks are reduced, which compensates
the effect of excessive $\Omega_{\nu}h^2$. Actually, the best-fit value of $h = 0.504$ resides at the edge of the prior. One may expect that it will become even smaller if the lower limit of the prior is decreased, and therefore the best-fit $\Omega_{\nu}h^2$ becomes larger. Obviously this is in severe conflict with measurements using supernova observations, $H_0 = 73.8 \pm 2.4 \mathrm{km}\cdot\mathrm{s}^{-1}\cdot\mathrm{Mpc}^{-1}$\cite{Riess2011}.

To test the goodness of the fit, we computed the $\chi^2$ of the best-fit TeVeS model and found $\chi^2=8292.54$. For comparison, the $\Lambda$CDM best fit gives $\chi^2=7791.18$\cite{PlanckCollaboration2013ES}. The difference is $\Delta\chi^2 = 501.36$. This is strong evidence that the TeVeS model considered in this paper cannot explain current CMB measurements. The difference in the $\chi^2$s is especially impressing, considering that the degree of freedom in the TeVeS is significantly increased. The original TeVeS model with a sterile neutrino is then ruled out by CMB observations. In Fig.\ref{fig.cmb_best} we can see that the best fit TeVeS model cannot properly fit the high-$l$ CMB power spectrum from {\it Planck}. Considering that $\Omega_{\nu}h^2$ has an important influence on the high-$l$ CMB power spectrum and its large best-fit value, this may suggest that the sterile neutrino is not a satisfactory substitute of DM in TeVeS cosmology.

\section{Conclusions}\label{sec.conclude}

In this paper we have tested the TeVeS theory
with several cosmological observations.
We have extended the previous probe of the late time structure growth by measuring the ratio of different perturbations $E_G$ in [48] to the complementary observations by measuring the overall amplitude of perturbations, such as the velocity spectrum and the kSZ effect, respectively.
We have found that the  dispersion of the baryon peculiar velocity at $r<100\mathrm{Mpc}$ in the
TeVeS cosmology is usually larger than that in the $\Lambda$CDM model and $\langle v_{\mathrm{b}}^2
\rangle$ decays faster in the TeVeS when the scale increases. We have computed the
linear and nonlinear kSZ anisotropy power
spectrum in TeVeS theory by assuming $\tau=0.09$. The linear kSZ
power spectra are within upper limits
measured by SPT and ACT, although they are much lower than those of the
$\Lambda$CDM model. The nonlinear kSZ power spectrum in the TeVeS model is in tension with measurements of the SPT and ACT, despite the uncertainties in
our theoretical prediction.

We have extended our discussions to the scale
dependence of the evolution of large scale
structure. In TeVeS cosmology, we have shown that
the normalized growth rate $f\sigma_8(k)$ rises
with the increase of $k$ at $z=0$. This is
clearly in contrast to the scale-independent
growth at subhorizon scales in the $\Lambda$CDM
model. Although the predicted $f\sigma_8$ in
TeVeS theory is consistent with the current
measurement using 6dF data, we expect that the
distinct scale dependence of the growth rate in the TeVeS model can potentially serve as a
powerful probe in distinguishing the
TeVeS from GR in future observations.

Considering the available high precision data on the cosmic microwave background radiations,
we have studied the CMB power spectrum in a TeVeS universe containing a sterile
neutrino. Fitting to {\it Planck} 2013 data,  we noticed that for the TeVeS cosmology,
although the constrained optical depth at the
border of the 68\% C.L. can give the end of
the reionization marginally allowed by the
constraints on the reionization history of our
Universe, the best fit value of the optical depth
is extremely small, which indicates that the end
of reionization happened at $z = 1.2$. The
constraints for the abundance of the sterile
neutrino and the Hubble's constant read
$\Omega_{\nu}h^2 = 0.156_{-0.002}^{+0.003}$ and
$H_0 < 50.8 \mathrm{km}\cdot
s^{-1}\cdot\mathrm{Mpc}^{-1}$ at 68\% C.L. The obtained
Hubble parameter is much lower than the observed value from supernovae
measurements. This is
certainly not allowed and it clearly rules out the TeVeS model considered in this work. Furthermore, comparing the $\chi^2$ of the best-fit TeVeS model with that of $\Lambda$CDM, we find that the TeVeS model is significantly disfavored by CMB observations. Because of the large uncertainties in current observations, the statistical significance of measurements of the kSZ effect and growth rate is now much weaker than CMB observations such as {\it Planck}. Yet we expect, with the improvement of the accuracy in their measurements, they will be useful in constraining cosmological models and become good complementary probes to CMB measurements.

In conclusion, we have examined the late time structure growth in the TeVeS model and found tensions between the TeVeS and cosmological observations. The conflict is more obvious when the TeVeS model is confronted by the CMB observations from {\it Planck}. Although the current available observational data from large scale structure growth are not precise enough to put tight constraints on the TeVeS model, our theoretical discussions on the density growth rate and its scale dependence in the TeVeS model can demonstrate that these complementary observable quantities have prospective abilities to distinguish general relativity and modified gravity theories at cosmological scales. With upcoming precise measurements, more studies in this respect are called for.

\begin{acknowledgments}
We acknowledge financial support from National
Basic Research Program of China (973 Program
2013CB834900 \& 2015CB857001) and National Natural Science
Foundation of China.
\end{acknowledgments}

\bibliography{tvs}

\end{document}